\documentclass[sigconf]{acmart}

\usepackage{booktabs} 

\usepackage{booktabs} 
\usepackage{amsmath}
\usepackage{amssymb}
\DeclareMathOperator*{\argmax}{arg\,max}
\usepackage{algorithm}  
\usepackage{algpseudocode}  
\usepackage{amsmath}  
\usepackage{array}
\usepackage{bm}

\begin{document}

%
\title[Designing for Serendipity in a University Course Recommendation System]{Combating the Filter Bubble: Designing for Serendipity in a University Course Recommendation System}

%
\author{Zachary A. Pardos}

\affiliation{%
  \institution{University of California, Berkeley}
  }
 \email{pardos@berkeley.edu}
  
\author{Weijie Jiang}
\affiliation{%
  \institution{University of California, Berkeley \& Tsinghua University}
  }
\email{jiangwj@berkeley.edu}



%
\renewcommand{\shortauthors}{Pardos and Jiang}

%
\begin{abstract}
Collaborative filtering based algorithms, including Recurrent Neural Networks (RNN), tend towards predicting a perpetuation of past observed behavior. In a recommendation context, this can lead to an overly narrow set of suggestions lacking in serendipity and inadvertently placing the user in what is known as a "filter bubble." In this paper, we grapple with the issue of the filter bubble in the context of a course recommendation system in production at a public university. Most universities in the United States encourage students to explore developing interests while simultaneously advising them to adhere to course taking norms which progress them towards graduation. These competing objectives, and the stakes involved for students, make this context a particularly meaningful one for investigating real-world recommendation strategies. We introduce a novel modification to the skip-gram model applied to nine years of historic course enrollment sequences to learn course vector representations used to diversify recommendations based on similarity to a student's specified favorite course. This model, which we call multifactor2vec, is intended to improve the semantics of the primary token embedding by also learning embeddings of potentially conflated factors of the token (e.g., instructor). Our offline testing found this model improved accuracy and recall on our course similarity and analogy validation sets over a standard skip-gram. Incorporating course catalog description text resulted in further improvements. We compare the performance of these models to the system's existing RNN-based recommendations with a user study of undergraduates (N = 70) rating six characteristics of their course recommendations. Results of the user study show a dramatic lack of novelty in RNN recommendations, a consequence of the filter bubble, and depict the characteristic trade-offs that make serendipity difficult to achieve. 
\end{abstract}

%
%

%

%
\maketitle

\section{Introduction}
Among the institutional values of a liberal arts university is to expose students to a variety of perspectives expressed in courses across its various physical and intellectual schools of thought. Collaborative filtering based sequence prediction methods, in this environment, can provide personalized course recommendations based on temporal models of normative behavior \cite{pardos2019connectionist} but are not well suited for surfacing courses a student may find interesting but which have been relatively unexplored by those with similar course selections to them in the past. Therefore, a more diversity oriented model can serve as an appropriate compliment to recommendations made from collaborative based methods. This problem of training on the past without necessarily repeating it is an open problem in many collaborative filtering based recommendation contexts, particularly social networks, where, in the degenerate cases, users can get caught in "filter bubbles," or model-based user stereotypes, leading to a narrowing of item recommendation variety \cite{nguyen2014exploring,zhang2012auralist,kay2000stereotypes}.

We introduce a novel skip-gram model variant into a production recommender system at a public university designed to surface serendipitous course suggestions. We use the definition of serendipity as user perceived unexpectedness of result combined with successfulness \cite{shani2011evaluating}, which we define as a course recommendation the student expresses interest in enrolling in. At many universities, conceptually similar courses exist across departments but use widely differing disciplinary vernacular in their catalog descriptions, making them difficult for learners to search for and to realize their commonality. We propose that by tuning a vector representation of courses learned from nine years of enrollment sequences, we can capture enough implicit semantics of the courses to more abstractly, and accurately construe similarity. To encourage the embedding to learn features that may generalize across departments, our skip-gram variants simultaneously learns department (and instructor) embeddings. While more advanced attention-based text generation architectures exist \cite{vaswani2017attention}, we demonstrate that properties of the linear vector space produced by "shallow" networks are of utility to this recommendation task. Our recommendations are made with only a single explicit course preference given by the user, as opposed to the entire course selection history needed by session-based Recurrent Neural Network approaches \cite{hidasi2016parallel}. Single example, also known as "one-shot," generalization is borrowed from the vision community, which has pioneered approaches to extrapolating a category from a single labeled example \cite{fei2006one,vinyals2016matching}. 

Other related work applying skip-grams to non-linguistic data include node embeddings learned from sequences of random walks of graphs \cite{ribeiro2017struc2vec} and product embeddings learned from ecommerce clickstream  \cite{chen2018behavior2vec}. Our work, methodologically, adds rigor to this approach by tuning the model against validation sets created from institutional knowledge and curated by the university. We conduct a user study (N = 70) of undergraduates at the university to evaluate their personalized course recommendations made by models designed for serendipity and by the RNN-based recommendations, which previously existed in the system. The findings underscore the tension between unexpectedness and successfulness and show the superiority of the skip-gram based method, as well as a bag-of-words baseline, for producing serendipitous results. From the open-response feedback received from students, we determined that the RNN-based recommendations still had a role to play, not in course exploration, but as a normative sorting of the order in which similar students had satisfied course requirements.

\section{Related Work}
In Natural Language Processing, a classical representation of words is as a vector of the contexts they appear in, equivalent to a word2vec model \citet{mikolov2013distributed} without its hidden layer. Such vector representations are called explicit, as each dimension directly corresponds to a particular context \cite{levy2014linguistic}. These explicit vector-space representations have been extensively studied in the NLP literature \cite{turney2010frequency, baroni2010distributional}, and are known to exhibit a large amount of attributional similarity \cite{pereira1993distributional, lin1998information, kotlerman2010directional}. Although \citet{baroni2014don} show that the neural embeddings obtain a substantial improvement against explicit representations on a wide range of lexical semantics tasks, \citet{levy2014linguistic} argue that under certain conditions traditional word similarities induced by explicit representations can perform just as well on lexical relationship validation sets. Their debates encourage us to utilize course descriptions to generate explicit bag-of-words representations for courses and compare them to our neural embedding models.

\citet{nguyen2014exploring} measured the "filter bubble" effect in terms of content diversity at the individual level and found that collaborative filtering-based recommender systems expose users to a slightly narrowing set of items over time. \citet{mcnee2006being} also proposed that the recommender community should move beyond  conventional accuracy metrics and their associated experimental methodologies. To counter the "filter bubble", \citet{zhang2012auralist} used a collection of novel LDA-based algorithms inspired by principles of "serendipitous discovery" and injected serendipity, novelty, and diversity to music recommendations whilst limiting the impact on accuracy. Different serendipitous metrics that measure the uncertainty and relevance of user's attitude towards items in order to mitigate the over-specialization problem with surprising suggestions are combined with traditional collaborative filtering recommendation \cite{pandey2018recommending} and content-based recommendation \cite{abbassi2009getting}. \citet{kawamae2009personalized} presented the Personal Innovator Degree (PID) which focused on the dynamics and precedence of user preference to recommend items that match the latest preference of the target user to achieve serendipity. 

Recommender systems in higher education contexts have recently focused on prediction of which courses a student will take or the grade they will receive if enrolled. At Stanford, a system called "CARTA" allows students to see grade distributions, course evaluations, and the most common courses taken before a course of interest \cite{chaturapruek2018data}. At UC Berkeley, our AskOski\footnote{https://askoski.berkeley.edu} recommender, named after the school's mascot, serves students next-semester course considerations based on their personal course enrollment history \cite{pardos2019connectionist}. Earlier systems included a focus on requirement satisfaction \cite{parameswaran2011recommendation} and career-based relevancy recommendation \cite{farzan2011encouraging}. No system has yet focused on serendipitous or novel course discovery.  


\section{models and methodology}
This section introduces three competing models used to generate our representations. The first model uses the \textit{skip-gram} model, which we refer to as \textit{course2vec} in this context, to learn course representations from enrollment sequences. Our second model is a novel variant on the skip-gram, introduced in this paper, which learns representations of explicitly defined features of a course (such as the instructor or department) in addition to the course representation. The intuition behind this approach is that the course representation could have, conflated in it, the influence of the multiple-instructors that have taught the course over the years. In order to separate out the department material of the course from the instructor, we allow for an instructor representation to be learned at the same time, but separate from the course representation. We contend that this may increase the fidelity of the course representation and serve as a more accurate representation of the essence of the course. The last representation model is a standard bag-of-words, constructed from course descriptions. In the last subsection, we describe the algorithm used to surface serendipitous recommendations using these course representations.
\subsection{Course2vec}
The \textit{course2vec} model involves learning distributed representations of courses from students' enrollment records throughout semesters by using a notion of a enrollment sequence as a "sentence" and courses within the sequence as "words", borrowing terminology from the linguistic domain. For each student \textit{s}, his/her chronological course enrollment sequence is produced by first sorting by semester then randomly serializing  within-semester course order. Then, each course enrollment sequence is trained on like a sentence in the Skip-gram model. The formulation of the \textit{course2vec} model is presented in our supplement (section 10.1). 


In language models, two word vectors will be cosine similar if they share similar sentence contexts. Likewise, in the university domain, courses that share similar co-enrollments, and similar previous and next semester enrollments, will likely be close to one another in the vector space. Course2vec learns course representations using a skip-gram model \cite{mikolov2013distributed} by maximizing the objective function over all the students' course enrollment sequences.

\subsection{Multi-factor Course2vec}
\begin{figure}
\includegraphics[height=2.3in, width=3in]{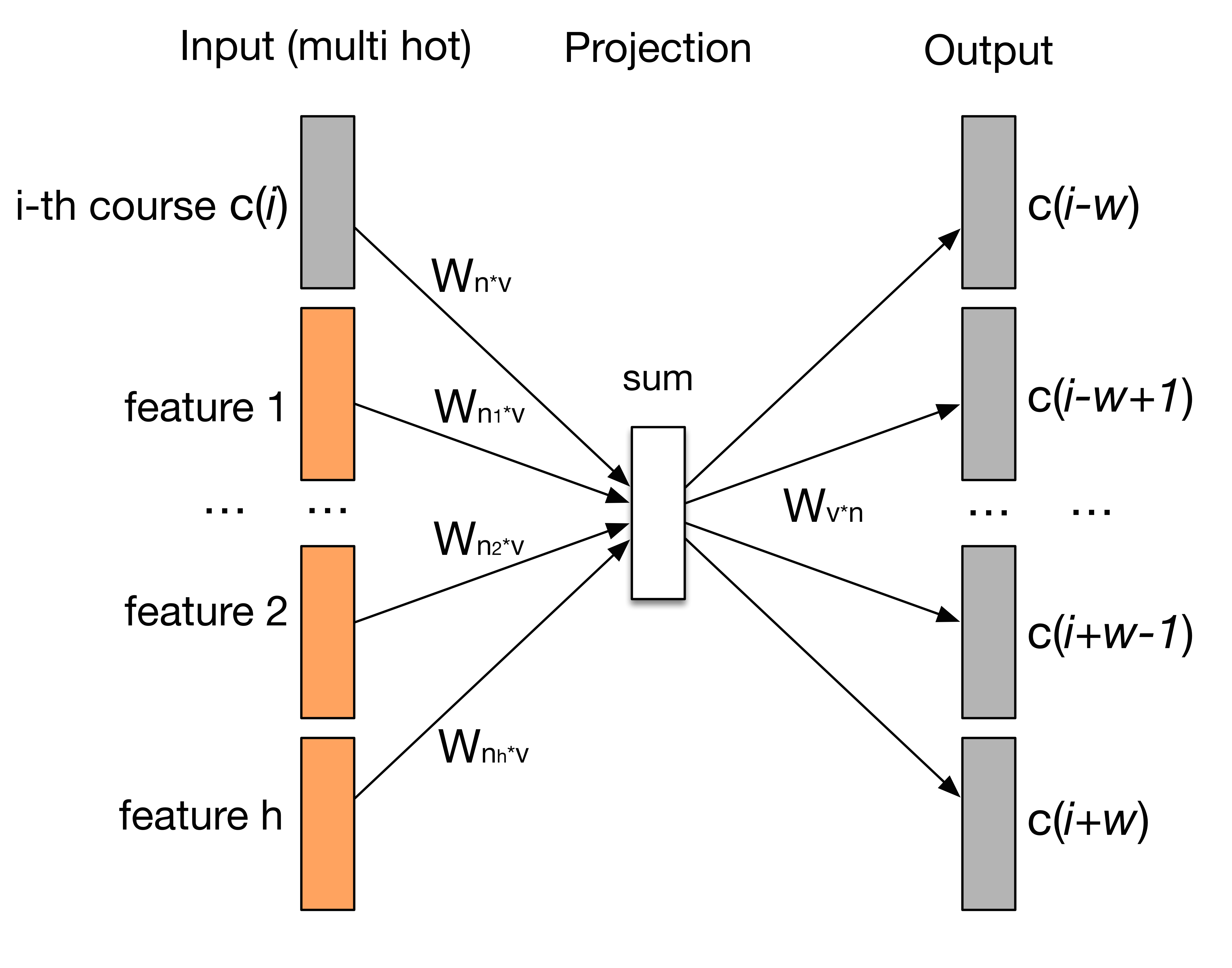}
\caption{multi-factor course2vec model}
\label{multi-c2v}
\end{figure}

The training objective of the skip-gram model is to find word representations that are useful for predicting the surrounding words in a sentence or a document. Each word in the corpus is used as an input to a log-linear classifier with continuous projection layer, to predict words within a certain range before and after the current word. Therefore, the skip-gram model can be also deemed as a classifier with input as a target course and output as a context course. In this section, we consider adding more features of courses to the input to enhance the classifier and its representations, as shown in Figure \ref{multi-c2v}.
Each course is taught by one or several instructors over the years and is associated with an academic department. 
The multi-factor course2vec model learns both course and feature representations by maximizing the objective function over all the students' enrollment sequences and the features of courses, defined as follows.
\begin{equation}
  \sum_{s\in S}\sum_{c_i\in s}\sum_{-w<j<w, j\neq 0}{\rm log} p(c_{i+j}|c_i, f_{i1}, f_{i2},..., f_{ih})
\label{loss-m}
\end{equation}
Probability $p(c_{i+j}|c_i, f_{i1}, f_{i2},..., f_{ih})$ of observing a neighboring course $c_{i+j}$ given the current course $c_i$ and its features $f_{i1}, f_{i2},..., f_{ih}$ can also be defined via the softmax function,
\begin{equation}
	p(c_{i+j}|c_i) = \frac{{\rm exp}(\boldsymbol{a}_i^T \boldsymbol{v}_{i+j}')}{\sum_{k=1}^n {\rm exp}(\boldsymbol{a}_i^T \boldsymbol{v}_k')}
  \label{softmax-m}
\end{equation}

\begin{equation}
	\boldsymbol{a}_i = \boldsymbol{v}_i+\sum_{j=1}^h \boldsymbol{W}_{n_j\times v}\boldsymbol{f}_{ij}
\end{equation}
where $\boldsymbol{a}_c$ is the vector sum of input course vector representation $\boldsymbol{v}_c$ and all the features vector representations of that course, $\boldsymbol{f}_{ij}$ is the multi-hot input of the j-th feature of course $i$, and $\boldsymbol{W}_{n_j \times v}$ is the weight matrix for feature $j$. So by multiplying $\boldsymbol{W}_{n_j \times v}$ and $\boldsymbol{f}_{ij}$, it gets the sum of feature vector representations of the i-th course. Here we keep the vector dimensions the same for both courses and features so that they are learned and mapped to the same vector space. 

\subsection{Bag-of-Words} 
A simple but indelible approach to item representation has been to create a vector, the length of the number of unique words across all items, with a non-zero value if the word in the vocabulary appears in it. Only unigram words are used to create this unordered vector list of words used to represent the document \cite{christopher2008introduction}.

The basic methodology based on bag-of words proposed by IR researchers for text corpora - a methodology successfully deployed in modern Internet search engines - reduces each document in the corpus to a vector of real numbers, each of which represents a term weight. The term weight might be:

\begin{itemize}
  \item a term frequency value indicating how many times the term occurred in the document.
  \item a binary value with 1 indicating that the term occurred in the document, and 0 indicating that it did not.
  \item tf-idf scheme \cite{dillon1983introduction}, the product of term frequency and inverse document frequency, which increases proportionally to the number of times a word appears in the document and is offset by the frequency of the word in the corpus and helps to adjust for the fact that some words appear more frequently in general.
\end{itemize}
We evaluate all three variants in our quantitative validation testing.

\subsection{Surfacing Serendipitous Recommendations from Course Representations}

We surface recommendations intended to be interesting but unexpected by finding an objective course $c_j$ which is most similar to a student's favorite course $c_i$ but diversifying the results by allowing only one result per department $d_j$:
\begin{equation}
c_j^* = \mathop{\argmax}_{c, d(c)=d_j} cos(c, c_i)
\end{equation}
where $d(c)$ means the the department of course $c$. Then all the counterpart courses $c_j^*$ in all the other departments will be ranked according to $cos(c_j^*, c_i)$, where $j=1,2...,k$. We can apply both neural representations and bag-of-words representations of courses in this method to generate the most similar courses in each department. 

\section{Experimental Environments}
\subsection{Off-line Dataset}
We used a dataset containing anonymous student course enrollments at UC Berkeley from Fall 2008 through Fall 2017. The dataset consists of per-semester course enrollment records for 164,196 students (both undergraduates and graduates) with a total of 4.8 million enrollments. A course enrollment record means that the student was still enrolled in the course at the end of the semester. Students at this university are allowed to drop courses up until close to the end of the semester without penalty. The median course load during students' active semesters was four. There were 9,478 unique lecture courses from 214 departments\footnote{At UC Berkeley, the smallest academic unit is called a "subject." For the purpose of communicability, we instead refer to subjects as departments in this paper.}  hosted in 17 different Divisions of 6 different Colleges. Course meta-information contains course number, department name, total enrollment and max capacity. In this paper, we only consider lecture courses with at least 20 enrollments total over the 9-year period, leaving 7,487 courses. Although courses can be categorized as undergraduate courses and graduate courses, all the students are allowed to enroll in many of the graduate courses no matter  their status. 
Enrollment data were sourced from the campus enterprise data warehouse with course descriptions sourced from the official campus course catalog API. We pre-processed the course description data in the following steps: (1) removing generic, often-seen, sentences across descriptions (2) removing stop words (3) removing punctuation (4) word lemmatization and stemming, and finally tokenizing the bag-of-words in each course description. We then compile the term frequency vector, binary value vector, and tf-idf vector for each course.

\subsubsection{Semantic Validation Sets}
In order to quantitatively evaluate how accurate the vector models are, a source of ground truth on the relationships between courses needed to brought to bear to see the degree to which the vector representations encoded this information. We used two such sources of ground truth to serve as validation sets, one providing information on similarity, the other on a variety of semantic relationships.

\begin{itemize}
\item \textit{Equivalency validation set}: A set of 1,351 course credit-equivalency pairs maintained by the Office of the Registrar were used for similarity based ground truth. A course is paired with another course in this set if a student can only receive credit for taking one of the courses. For example, an honors and non-honors version of a course will be appear as a pair because faculty have deemed that there is too much overlapping material between the two for a student to receive credit for both. 
\item \textit{Analogy validation set}: The standard method for validating learned word vector has been to use analogy to test the degree to which the embedding structure contains semantic and syntactic relationships constructed from prior knowledge. In the domain of university courses, we use course relationship pairs constructed from prior work using first-hand knowledge of the courses \cite{pardos2018map}. The 77 relationship pairs were in five categories; online, honors, mathematical rigor, 2-department topics, and 3-department topics. An example of an "online" course pair would be Engineering 7 and its online counterpart, Engineering W7 or Education 161 and W161. An analogy involving these two paris could be calculated as; Engineering 7W - Engineering 7 + Education 161 = Education W161.
\end{itemize}

\subsection{Online Environment (System Overview)}
The production recommender system at UC Berkeley uses a student data pipeline with the enterprise data warehouse to keep up-to-date enrollment histories of students. Upon CAS login, these histories are associated with the student and passed through an RNN model, which cross-references the output recommendations with the courses offered in the target semester. Class availability information is retrieved during the previous semester from a campus API once the registrar has released the schedule. The system is written with an AngularJS front-end and python back-end service which loads the machine learned models written in pyTorch. These models are version controlled on github and refreshed three times per semester after student enrollment status refreshes from the pipeline. The system receives traffic of around 20\% of the undergraduate student body from the UC Berkeley Registrar's website.

\section{Vector Model Refinement Experiments}
In this section, we first introduce our experiment parameters and the ways we validated the representations quantitatively. Then, we describe various ways we refined the models and the results of these refinement. 

\subsection{Model Evaluations}
We trained the models described in Section 3 on the students enrollment records data set. Specifically, we added the instructor(s) who teach the course and the course department as two features of courses in the multi-factor course2vec model. To ensure reproducibility, we put our model experiment settings to our supplement (section 10.2).

To evaluate course vectors on the course equivalency validation set, we fixed the first course in each pair and rank all the other courses according to their cosine similarity to the first course in descending order. We then noted the rank of the expected second course in the pair and describe the performance of each model on all validation pairs in terms of of \textit{mean rank}, \textit{median rank} and \textit{recall@10}.

For evaluation of the course analogy validation set, we followed the paradigm of analogy: $course2 - course1 + course3 \approx course4$. Courses were ranked by their cosine similarity to $course2 - course1 + course3$. An analogy completion is considered accurate (a hit) if the first ranked course is the expected $course4$ (excluding the other three from the list). We calculated the average accuracy (recall@1) and the recall@10 over all the analogies in the analogy validation set.

\subsection{Course2vec vs. Multi-factor Course2vec}
In this section, we first compared the pure course2vec model with the course representations from the multi-factor course2vec model using instructor, department, and both as factors. To further explore improvements to performance, we concatenated the primary course representational layer ($W_{n\times v}$ in Figure \ref{multi-c2v}) with the output representation layer ($W_{v\times n}'$ in Figure \ref{multi-c2v}), as demonstrated to be effective in the language domain \cite{gabor2017exploring}. 


Results of evaluation on the equivalency validation are shown in Table \ref{t-c2v-e} with analogy validation results shown in Table \ref{t-c2v-a}. Models using the output representation concatenation are labeled with "(+out)" next to their names in both tables. The multi-factor model outperformed the pure course2vec model in terms of recall@10 in both validation sets, with the combined instructor and department factor model performing the best. The same result held for the metrics of mean and median rank in equivalency, but multi-factor models were not always the best in terms of analogy recall@1 (Accuracy). Output layer concatenation did not improve any of the models on the equivalency validation but, interesting, improved all but the instructor model (recall@10) in the analogy validation set. The best recall achieved in analogies was by the instructor and department multi-factor model, successfully completing 85.57\% of course analogies when considering the top 10 candidates of each analogy completion.

\begin{table}
\small
  \caption{Equivalency validation of all the models}
  \label{t-c2v-e}
  \begin{tabular}{ >{\centering\arraybackslash}p{4.6cm} >{\centering\arraybackslash}p{2cm} >{\centering\arraybackslash}p{0.8cm}}
    \toprule
    \centering
    Model&Mean / Median Rank & Recall @10\\
    \midrule
    \centering
    course2vec& 244 / 21&0.3839\\
    course2vec (+out)&270 / 26 & 0.3430\\
    ins-course2vec&302 / 16&0.4406\\
    ins-course2vec (+out)& 400 / 32&0.3478\\
    dept-course2vec& 261 / 17&0.4005\\
    dept-course2vec (+out)&306 / 19&0.3721\\
    ins-dept-course2vec&224 / \textbf{15}&\textbf{0.4485}\\
    ins-dept-course2vec (+out)&\textbf{201} / 16&0.4312\\
    \hline
    tf & 589 / 5& 0.5451\\
    binary&612  / 6& 0.5308\\
    tf-idf& \textbf{566} / \textbf{4}& \textbf{0.5647}\\
    \hline
    tf+insdept-course2vec &168 / 6& 0.5691\\
    tf+insdept-course2vec (norm)&132 / \textbf{3}&0.6371\\
    bin.+insdept-course2vec&178 / 7& 0.5404\\
    bin.+insdept-course2vec (norm)&\textbf{129} / \textbf{3}&0.6251\\
    tfidf+insdept-course2vec& 213 / 14&0.4428\\
    \textbf{tfidf+insdept-course2vec (norm)}&132 / \textbf{3}&\textbf{0.6435}\\
  \bottomrule
\end{tabular}
\end{table}

\begin{table}
\small
\centering
  \caption{Analogy validation of all the models}
  \label{t-c2v-a}
  \begin{tabular}{cccl}
    \toprule
    Model&Accuracy&Recall@10\\
    \midrule
    course2vec&0.4739& 0.7539\\
    course2vec (+out)& 0.5011&0.7685\\
   
    ins-course2vec&0.5025&0.8094\\
    ins-course2vec (+out)&\textbf{0.5138}&0.7853\\
   
    dept-course2vec&0.3504&0.8257\\
    dept-course2vec (+out)&0.3581&0.8284\\

    ins-dept-course2vec&0.4784&0.8434\\
    ins-dept-course2vec (+out)&0.4961&\textbf{0.8557}&\\
    
    \hline
    tf & 0.3037&0.537\\
    binary&0.3159&\textbf{0.581}\\
    tf-idf& \textbf{0.3227}&0.542\\
    
    \hline
     tf+insdept-course2vec &0.5066&0.8438\\
   tf+insdept-course2vec (norm)&0.448&0.6872\\

    \textbf{bin.+insdept-course2vec}&\textbf{0.5193}& \textbf{0.8788}\\
    bin.+insdept-course2vec (norm)&0.4603&0.7449\\

    tfidf+insdept-course2vec&0.5138&0.8584\\
    tfidf+insdept-course2vec (norm)&0.4503&0.7059\\
  \bottomrule
\end{tabular}
\end{table}


For results in the following sections, we will use the non output concatenation versions of course2vec for equivalency validation comparisons and the concatenation versions for analogies validation comparison.

\subsection{Bag-of-words vs. Multi-factor Course2vec}
Among the three bag-of-words models, tf-idf performs the best in all equivalency set metrics, as seen in Table \ref{t-c2v-e}. The median rank (best=4) and recall@10 (best=0.5647) for the bag-of-words models were also substantially better than the best course2vec models, which had a best median rank of 15 with best recall@10 of 0.4485 for the instructor and department model. All course2vec models; however, showed better mean rank performance (best=224) compared with bag-of-words (best=566). This suggests that there are many outliers where literal semantic similarity (bag-of-words) is very poor at identifying equivalent pairs, whereas course2vec has much fewer near worst-case examples. This result is consistent with prior work comparing pure course2vec models to binary bag-of-words \cite{pardos2019connectionist}.

When considering performance on the analogy validation (Table \ref{t-c2v-a}), the roles are reversed, with all course2vec models performing better than the bag-of-words models in both accuracy and recall@10. The difference in recall of bag-of-words compared to course2vec when it comes to analogies is substantially greater (0.581 vs 0.8557), than the superiority difference of bag-of-words over course2vec in the equivalency validation (0.5647 vs 0.4485). 
These analyses establish that bag-of-words models are supreme in capturing course similarity, but are substantially inferior to our skip-gram based models in the more complex task of analogy completion. 

The comparison of course2vec related models and bag-of-words models on equivalency validation and analogy validation is to some extent counter to \citet{levy2014linguistic}'s argument that \citet{mikolov2013distributed}'s word analogy exploring method of first adding and subtracting word vectors, and then searching for a word similar to the result, is equivalent to searching for a word that maximizes a linear combination of three pairwise word similarities, instead of vector offsets encoding relational semantics. Otherwise, bag-of-words representations should also performs better on course analogies if the calculation on finding analogies is also based on cosine similarities. 
On top of that, all the neural embeddings perform better than bag-of-words representations on both accuracy and recall@10, which surfaces the signal that there is also relational semantics conveyed from course enrollment behaviors but not encoded in course semantic descriptions.

\subsection{Combining Bag-of-words and Course2vec Representations}
In light of the strong analogy performance of course2vec and strong equivalency performance bag-of-words in the previous section, we use the neural embeddings learned by mult-factor course2vec which incorporates both instructor and department to concatenate with bag-of-words representations. To counterbalance the different magnitudes of neural embeddings and bag-of-words representations, we create a normalized version of each vector set for comparison to non-normalized sets. 

We find that the normalized concatenation performs substantially better on the equivalency test than the previous best model in terms of recall@10 (0.6435 vs. 0.5647) as seen in Table \ref{t-c2v-e}. While the median rank of the concatenated model only improved one rank, from 4 to 3, the mean rank improved dramatically (from 566 to 132), and is the best of all models tested in terms of mean rank. Non-normalized vectors did not show improvements over bag-of-words alone in median rank and recall@10. Improvements in the analogy test (Table \ref{t-c2v-a}) were much more mild, with a recall@10 of 0.8788 of the best concatenated model vs. 0.8557 of the best course2vec only model. Normalization in the case of analogies, hurt all model performance, the opposite of what was observed in the equivalency test. This suggests that normalization improves local similarity but acts to degrade the more global structure of the vector space.

\section{User Study}
A user study was conducted to evaluate differences in recommendations among our developed representation based recommendation algorithms along five dimensions of quality. Students were asked to rate course recommendations in terms of their (1) unexpectedness (2) successfulness - or interest in taking the course (3) novelty (4) diversity of the results (5) and identifiable commonality among the results. \citet{shani2011evaluating} define serendipity as the combination of "unexpectedness" and "success." In the case of a song recommender, for example, success would be defined as the user listening to the recommendation. In our case, we use a student's expression of interest in taking the course as a proxy for success. The mean of their unexpectedness and successfulness rating will comprise our measure of serendipity. We evaluated three of our developed models, all of which displayed 10 results, only showing one course per department in order to increase diversity (and unexpectedness). The models were (1) the best \textit{BOW} model (tf-idf), (2) the best \textit{Analogy} validation model (bin.+insdept-course2vec), (3) and the best \textit{Equivalency} validation model (tfidf+insdept-course2vec norm). To measure the impact our department diversification filter would have on serendipity, we added a version of the best Equivalency model that did not impose this filter, allowing multiple courses to be displayed from the same department if they were the most cosine similar to the user's specified favorite course. Our fifth comparison recommendation algorithm was a collaborative-filtering based Recurrent Neural Network (RNN) that recommends courses based on a prediction of what the student is likely to take next given their personal course history and what other students with a similar history have taken in the past \cite{pardos2019connectionist}. We put a brief summary of the collaborative-filtering based Recurrent Neural Network (RNN) recommendation algorithm to our supplement (section 10.3). All five algorithms were integrated into a real-time recommender system for the purpose of this study and evaluated by 70  undergraduates at the university.
\subsection{Study Design}
Undergraduates were recruited from popular university associated Facebook groups and asked to sign-up for a one hour evaluation session. Since they would need to specify a favorite course they had taken, we restricted participants to those who had been at the university at least one full semester and were currently enrolled. The study was run at the beginning of the Fall semester, while courses could still be added and dropped and some students were still shopping for courses. We used a within-departments design whereby each volunteer rated ten course recommendations made by each of the five algorithms. Because of the considerable number of ratings expected ([3*10+2]*5 = 160) and the importance for students to carefully consider each recommended course, in-person sessions were decided on over asynchronous remote in order to better encourage on-task behavior throughout the session. Student evaluators were compensated with a \$40 gift card to attend one of four sessions offered across three days with a maximum occupancy of 25 each session. A total of 70\footnote{Due to an authentication bug during the fourth session, all twenty participating students were not able to access the collaborative recommendations of the fifth algorithm. RNN results in the subsequent section are therefore based on the 50 students from the first three sessions. When paired t-tests are conducted between RNN and the ratings of other algorithms, the tests are between ratings among these 50 students.} students participated. 

We began the session by introducing the evaluation motivation as a means for students to help inform the choice of algorithm that we will use for a future campus-wide deployment of a course exploration tool. Students started the evaluation by visiting a survey URL that asked them to specify a favorite course they had taken at the university. This favorite course was used by the first four algorithms to produce 10 course recommendations each, which included the course's department, course number, title, and full catalog description. There was a survey page for each algorithm in which students were asked to read the course descriptions carefully and then  rate each of the ten courses individually for their five point Likert scale agreement with the following statements: (1) This course was unexpected (2) I am interested in taking this course (3) I did not know about this course before. These ratings respectively measured unexpectedness, successfulness, and novelty. After rating the individual courses, students were asked to rate their agreement with the following statements pertaining to the results as a whole: (1) Overall, the course results were diverse (2) The course results shared something in common with my favorite course. These ratings measured dimensions of diversity and commonality. Lastly, students were asked to provide an optional follow-up open text response to the question, "If you identified something in common with your favorite course, please explain it here." On the last page of the survey, students were asked to specify their major, year, and to give optional open response feedback on their experience. Graduate courses were not included in the recommendations and the recommendations were not limited to courses available in the current semester.
\begin{table*}
\small
\caption{Average student ratings of recommendations from the five algorithms across the six measurement categories.}
\label{main-results}
\begin{tabular}{lllllll}
\toprule
algorithm & unexpectedness & successfulness & serendipity & novelty & diversity & commonality \\
\midrule
BOW (div) & \textbf{3.550} & 2.904 & \textbf{3.227} & \textbf{3.896} & 4.229 & 3.229 \\
Analogy (div) & 3.473 & 2.851 & 3.162 & 3.310 & \textbf{4.286} & 2.986 \\
Equivalency (div) & 3.297 & 2.999 & 3.148 & 3.323 & 4.214 & 3.257 \\
Equivalency (non-div) & 2.091 & \textbf{3.619} & 2.855 & 2.559 & 2.457 & \textbf{4.500} \\
RNN (non-div) & 2.184 & 3.566 & 2.875 & 1.824 & 3.160 & 4.140\\
\bottomrule
\end{tabular}
\end{table*}
\subsection{Results}
Results of average student ratings of the five algorithms across the six measurement categories are shown in Table \ref{main-results}. The diversity based algorithms, denoted by "(div)," all scored higher than the non-diversity (non-div) algorithms in unexpectedness, novelty, diversity, and the primary measure of serendipity. The two non-diversity based algorithms; however, both scored higher than the other three algorithms in successfulness and commonality. All pairwise differences between  diversity and non-diversity algorithms were statistically significant, using the p < 0.001 level after applying a Bonferoni correction for multiple (60) tests. Within the diversity algorithms, there were no statistically significant differences except for BOW scoring higher than Equivalency (div) on unexpectedness and scoring higher than both Equivalency (div) and Analogy (div) on novelty. Among the two non-diversity algorithms, there were no statistically significant differences except for the RNN scoring higher on diversity and Equivalency (non-div) recommendations scoring higher on novelty. With respect to measures of serendipity, the div and non-div algorithms had similar scores among their respective strengths (3.473-3.619); however, the non-div algorithms scored substantially lower in their weak category of unexpectedness (2.091 $\&$ 2.184) than did the div algorithms in their weak category of successfulness (2.851-2.999), resulting in statistically significantly higher serendipity scores for the div algorithms. 

\begin{figure}[t]
\includegraphics[height=1.7in, width=3in]{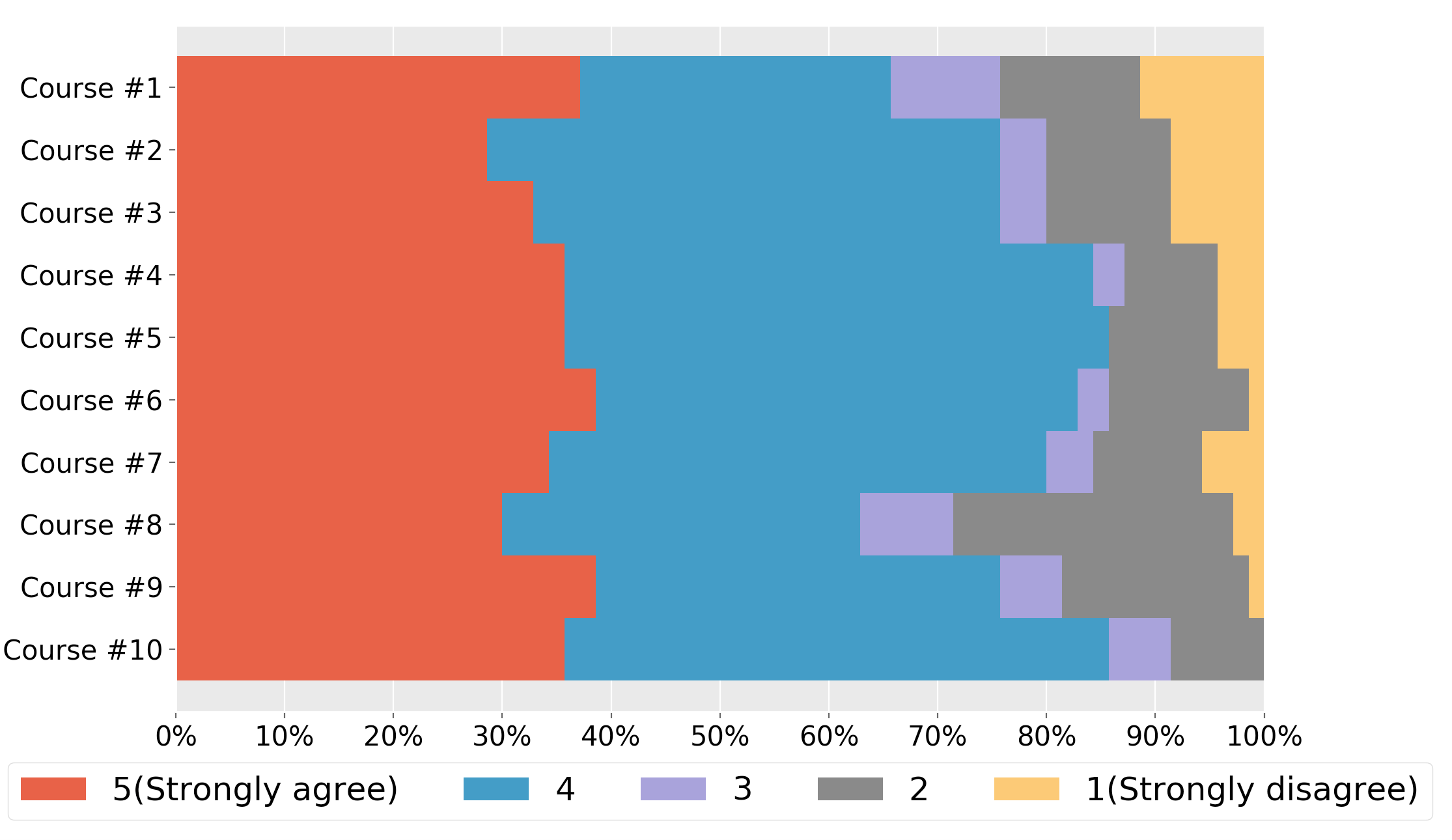}
\caption{Novelty rating proportions for BOW (div)}
\label{n-b}
\end{figure}

\begin{figure}
\includegraphics[height=1.7in, width=3in]{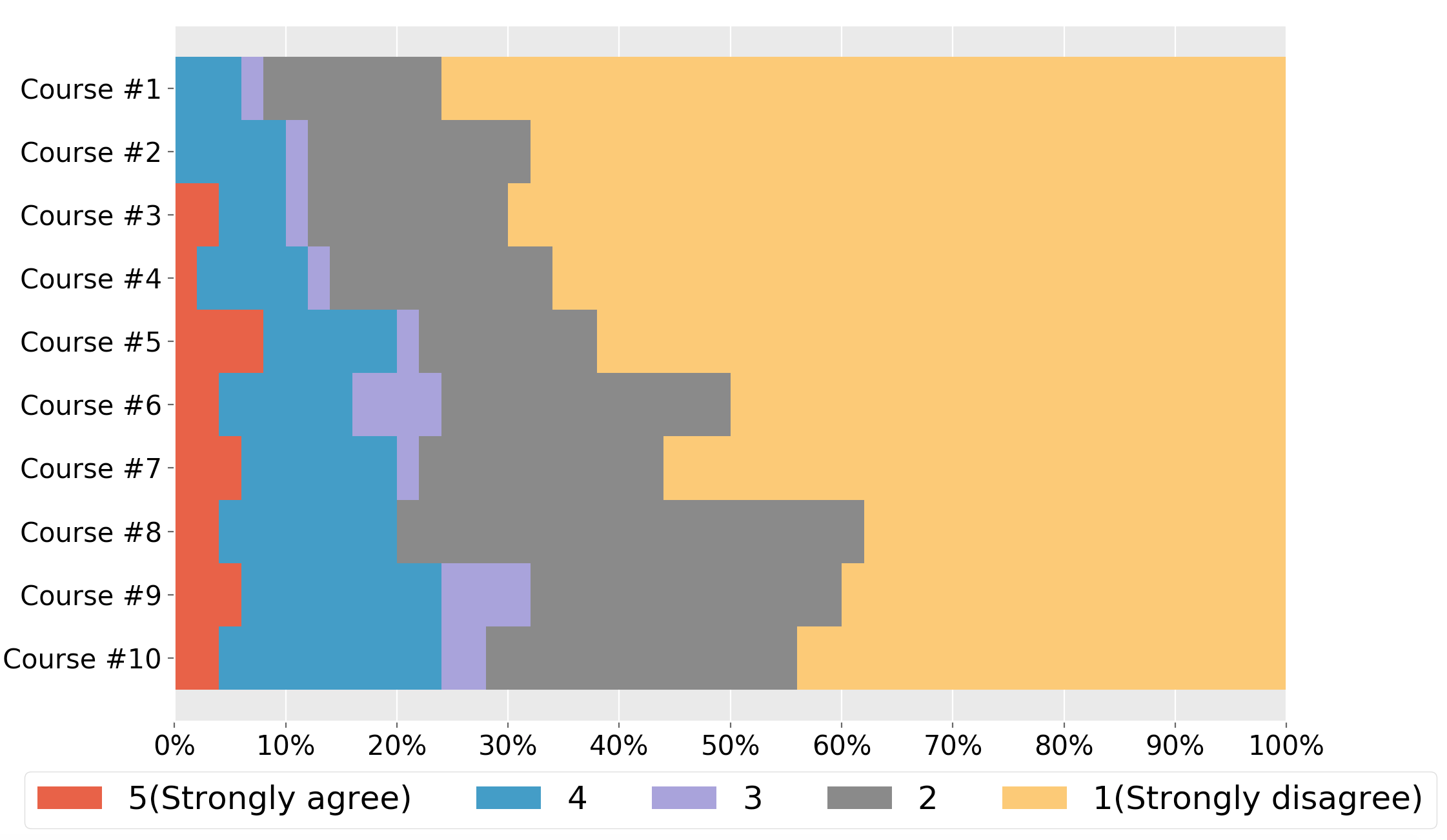}
\caption{Novelty rating proportions for RNN (non-div)}
\label{n-r}
\end{figure}

\begin{figure}
\includegraphics[height=2.2in, width=2.8in]{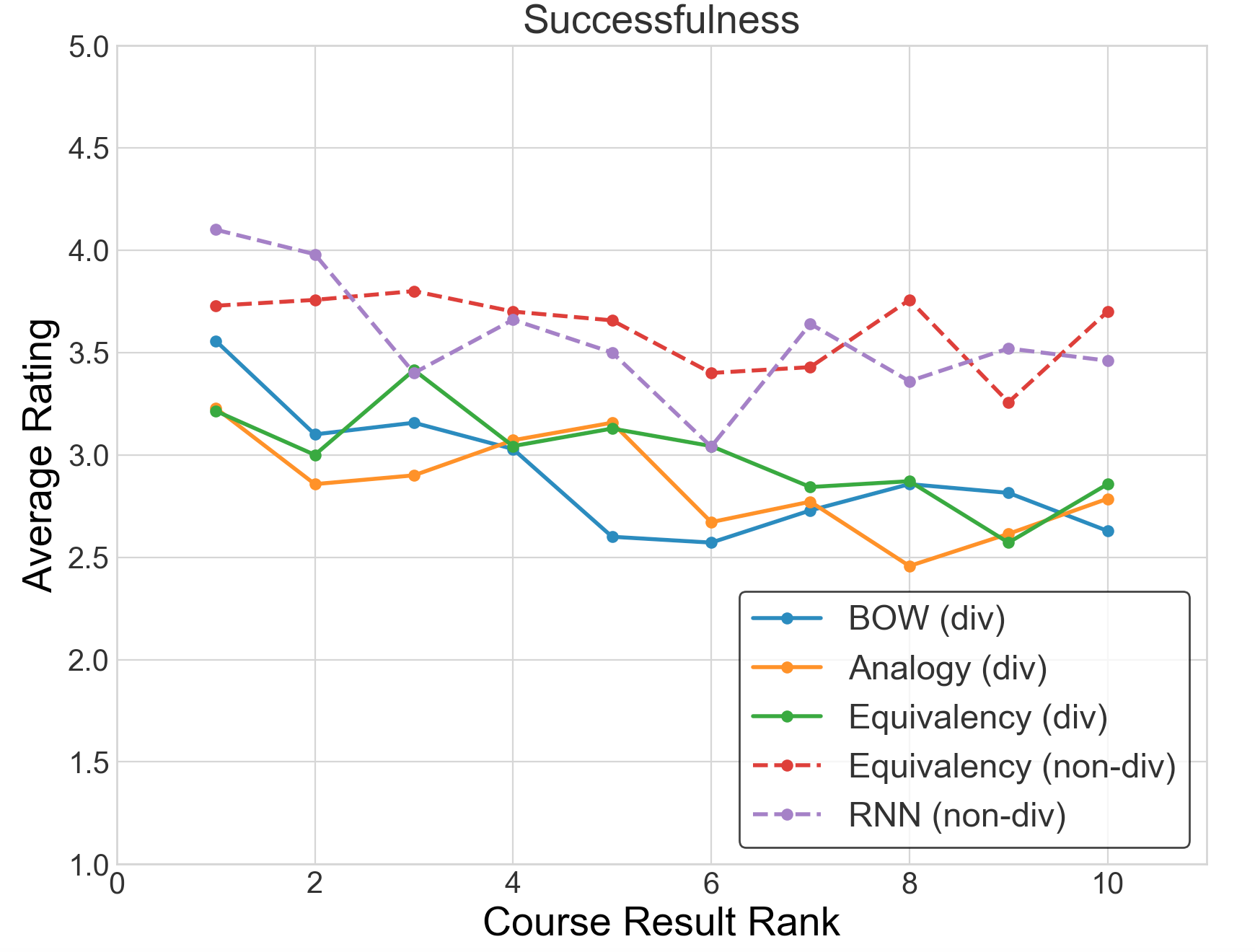}
\caption{Successfulness comparison}
\label{s-c}
\end{figure}

\begin{figure}
\includegraphics[height=1.6in, width=3in]{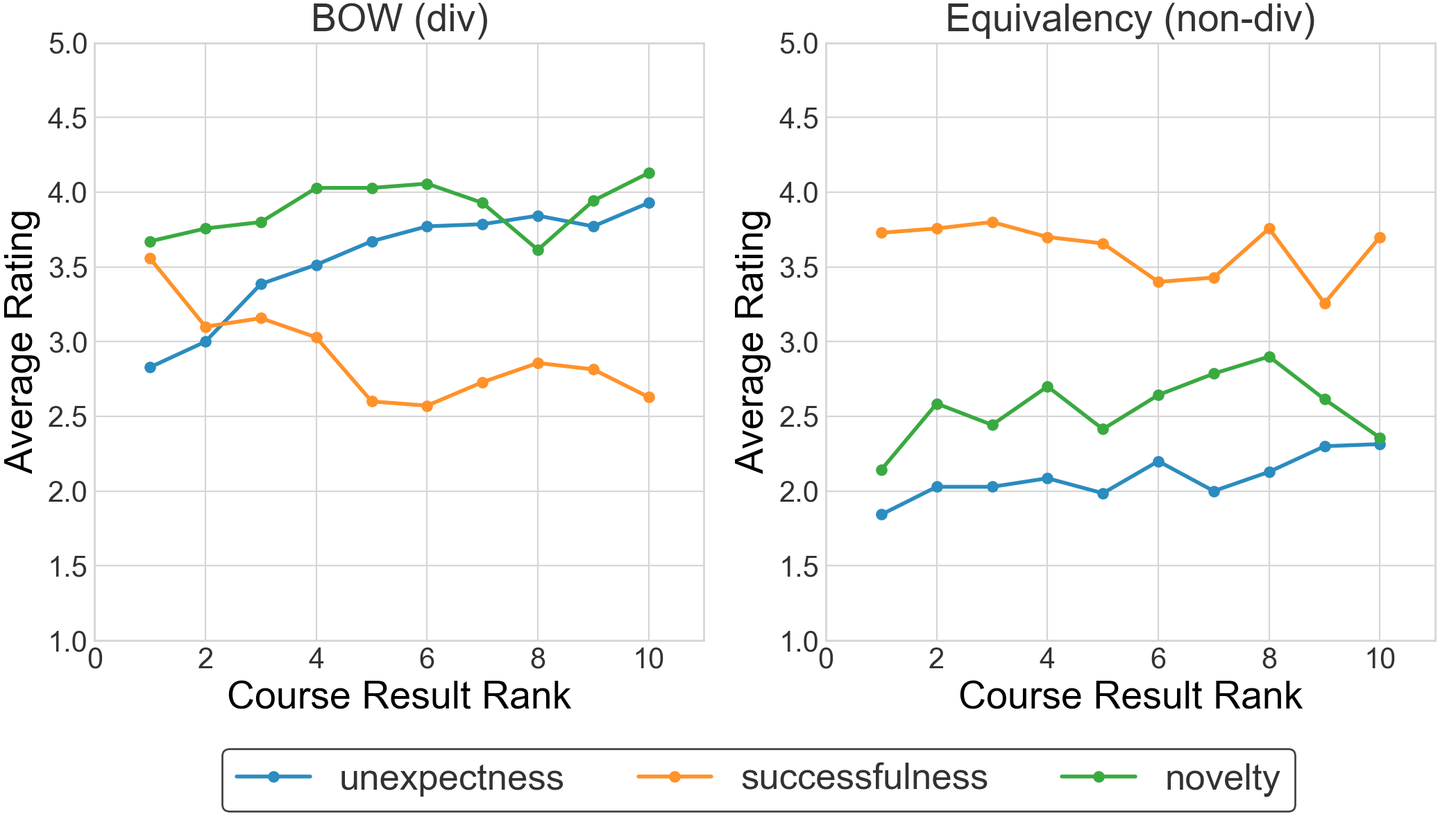}
\caption{BOW (div) vs. Equivalency (non-div) comparison}
\label{r-c}
\end{figure}

The most dramatic difference can be seen in the measure of novelty, where BOW (div) scored 3.896 and RNN (non-div) scored 1.824, the lowest rating in the results matrix. The proportion of each rating level given to the two algorithms on this question is shown in Figures \ref{n-b} and \ref{n-r}. Hypothetically, an algorithm that recommended randomly selected courses would score high in both novelty and unexpectedness, and thus it is critical to also weigh their ability to recommend courses that are also of interest to students. Figure \ref{s-c} shows successfulness ratings for each of the algorithms aggregated by rank of the course result. The non-div algorithms, shown with dotted lines, always perform as well or better than the non-div algorithms at every rank. The more steeply declining slope of the div algorithms depicts the increasing difficulty of finding courses of interest across different departments. The tension between the ability to recommend courses of interest that are also unexpected is shown in Figure \ref{r-c}, where BOW (div) is able to recommend courses of interest but low unexpectedness in the top few results. These values quickly swap places the lower the result ranks go. The non-diversity algorithms, on the other hand, maintain high successfulness but also low unexpectedness throughout the 10 result ranks.

Are more senior students less likely to rate courses as novel or unexpected, given they have been at the university longer and been exposed to more courses? Among our sophomore (27), junior (22), and senior (21) level students, there were no statistically significant trends among the six measures, except for a marginally significant trend (p = 0.007, shy of the p < 0.003 threshold given the Bonferroni correction) of more senior students rating recommendations as less unexpected (avg = 2.921) than juniors (avg = 3.024), whose ratings were not statistically separable from sophomores (avg = 3.073).

\subsection{Qualitative Characterization of Algorithms}
In this section, we attempt to synthesize qualitative characterizations of the different algorithms by looking at the open responses students gave to the question asking them to describe any commonalities they saw among recommendations made by each algorithm to their favorite course.
\subsubsection{BOW (div)}
Several students remarked positively about recommendations matching to the themes of "art, philosophy, and society" or "design" exhibited in their favorite course. The word "language" was mentioned by 14 of the 61 respondents answering the open response question. Most of these comments were negative, pointing out the limitations of similarity matching based solely on literal course description matching. The most common critique given in this category was of the foreign spoken language courses that showed up at the lower ranks when students specified a favorite course involving programming languages. Other students remarked at the same type of occurrence when specifying a favorite course related to cyber security, receiving financial security courses in the results.
\subsubsection{Analogy (div)}
The word "interesting" appeared in seven of the 54 comments left by students to describe commonalities among the analogy validation optimized algorithm. This word was not among the top 10 most frequent words in any of the other four algorithms. Several students identified broad themes among the courses that matched to their favorite course, such as "identity" and "societal development." On the other end of the spectrum, one student remarked that the results "felt weird" and were only "vaguely relevant." Another student stated that, "the most interesting suggestion was the Introduction to Embedded Systems [course] which is just different enough from my favorite course that it's interesting but not too different that I am not interested," which poignantly articulates the crux of difficulty in striking a balance between interest and unexpectedness to achieve a serendipitous recommendation.
\subsubsection{Equivalency (div)}
Many students (seven of the 55) remarked positively on the commonality of the results with themes of data exhibited by their favorite course (in most cases STATS C8, an introductory data science course). They mentioned how the courses all involved "interacting with data in different social, economic, and psychological contexts" and "data analysis with different applications." One student remarked on this algorithm's tendency to match at or around the main topic of the favorite course, further remarking that "they were relevant if looking for a class tangentially related." 
\subsubsection{Equivalency (non-div)}
This algorithm was the same as the above, except that it did not limit results to one course per department. Because of this lack of department filter, 15 of the 68 students submitting open text responses to the question of commonality pointed out that the courses returned were all from the same department. Since this model scored highest on a validation task of matching to a credit equivalent course pair (almost always in the same department), it is not surprising that students observed that results from this algorithm tended to all come from the department of the favorite course, which also put it close to their nexus of interest.
\subsubsection{RNN (non-div)}
The RNN scored lowest in novelty, significantly lower than the other non-div algorithm, and was not significantly different from the other non-div algorithm in successfulness. In this case, what is the possible utility of the collaborative-based RNN over the non-div Equivalency model? Many of the 47 (of 50) student answers to the open response commonality question point at a potential answer of major related (mentioned by 21 students) and courses that fulfilled a requirement (mentioned by seven) as the distinguishing signature of this algorithm. Since the RNN is based on normative next course enrollment behavior, it is reasonable that it suggested many courses that satisfy an unmet requirement. This algorithm's ability to predict student enrollments accurately became a detriment to some, as seven remarked that it was recommending courses that they were currently enrolled in. Due to the institutional data refresh schedule, student current enrollments are not known until after the add/drop deadline. This may be a shortcoming that can be rectified in the future.

\section{Feature re-design}
\begin{figure}[h]
\centering
\includegraphics[height=4.7in, width=3.3in]{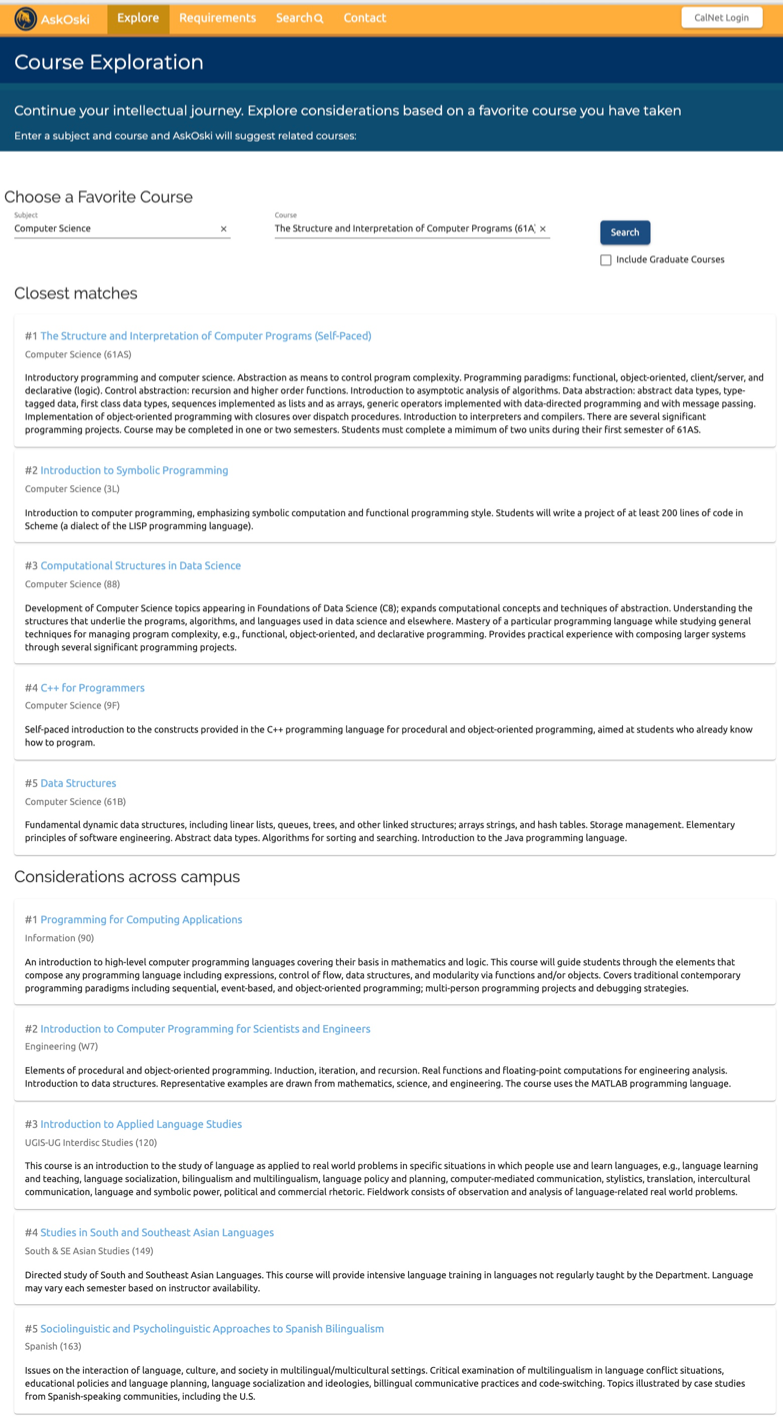}
\caption{The ``Explore" Interface}
\label{rnn}
\end{figure}

As a result of the feedback received from the user study, we worked with campus to pull down real-time information on student requirement satisfaction from the Academic Plan Review module of the PeopleSoft Student Information System. we re-framed the RNN feature as a "Requirements" satisfying feature that, upon log-in, shows students their personalized list of unsatisfied requirements. After selecting a requirement category to satisfy, the system displays  courses which satisfy the selected requirement and are offered in the target semester. The list of courses is sorted by the RNN to represent the probability that students like them will take the class. This provides a signal to the student of what the normative course taking behavior is in the context of requirement satisfaction. For serendipidous suggestions, we created a separate "Explore" tab (Figure \ref{rnn}) using the Equivalency (non-div) model to display results of the top five courses most similar within the same department, due to its strong successfulness ratings, and the BOW (div) model to surface the top five courses similar across departments, due to its strong serendipidous and novelty ratings.

\section{Discussion}
Surfacing courses that are of interest but not known before means expanding a student's knowledge and understanding of the University's offerings. As students are exposed to courses that veer further from their home department and nexus of interest and understanding, recommendations become less familiar with descriptions that are harder to connect with. This underscores the difficulty of producing an unexpected but interesting course suggestion, as it often must represent a recommendation of uncommon wisdom in order to extend outside of a student's zone of familiarity surrounding their centers of interest. Big data can be a vehicle for, at times, reaching that wisdom. Are recommendations useful when they suggest something expected or already known? Two distinct sets of responses to this question emerged from student answers to the last open ended feedback question. One representative remark stated, "the best algorithms were the ones that had more diverse options, while still staying true to the core function of the class I was searching. The algorithms that returned classes that were my major requirements/in the same department weren't as helpful because I already knew of their existence as electives I could be taking." While a different representative view was expressed with, "I think the fifth algorithm [RNN] was the best fit for me because my major is pretty standardized" These two comments make a case for both capabilities being of importance. They are also a reminder of the desire among young adults for the socio-technical systems of the university to offer a balance of information, exploration and , at times, guidance. 

\section{Limitations}
The more semantically distal, even if conceptually similar, the less a student may be able to recognize the commonality with a favorite course. A limitation of demonstrating the utility of the neural embeddings is that students had to rely on the course description semantics in order to familiarize themselves with the suggested course. If a concept was detected by the neural embeddings but not the BOW, this means it may have been difficult for students to pick-up on this from the descriptions alone. Future evaluations could include links to additional course information, such as course syllabi. While our diversity algorithms produced serendipitous recommendations, they will not always produce the desired recommendations with respect to interest if the most salient features of the course's embedding is not what the student most liked about it.

\bibliographystyle{ACM-Reference-Format}
\bibliography{sample-bibliography}

\end{document}